\documentclass[aps,longbibliography,
  twocolumn,amsmath,
  amssymb,nofootinbib, superscriptaddress, preprintnumbers]{revtex4-1}

\usepackage{graphics}      
\usepackage{graphicx}      
\usepackage{longtable}     
\usepackage{url}           
\usepackage{bm}            
\usepackage{xcolor}
\usepackage{mathrsfs}
\usepackage[colorlinks,allcolors=blue]{hyperref}
\usepackage{blindtext}

\DeclareFontFamily{OT1}{pzc}{}
\DeclareFontShape{OT1}{pzc}{m}{it}{<-> s * [1.100] pzcmi7t}{}
\DeclareMathAlphabet{\mathpzc}{OT1}{pzc}{m}{it}

\begin{document}

\preprint{MPP-2020-124}

\title{Turbulence Fingerprint on Collective Oscillations of Supernova Neutrinos}

\newcommand*{\redl}{\textcolor{red}{\{}}
\newcommand*{\redr}{\textcolor{red}{\}}}

\newcommand*{\MPP}{\textit{\small{Max-Planck-Institut f\"ur Physik (Werner-Heisenberg-Institut), F\"ohringer Ring 6, 80805 M\"unchen, Germany}}}
\author{Sajad Abbar\\  \MPP }

\begin{abstract}

We  bring to light a  \emph{novel} mechanism through which 
 turbulent matter density fluctuations  can induce collective neutrino flavor conversions in 
 core-collapse supernovae, i.e., the \emph{leakage} of flavor instabilities between different Fourier modes.
The leakage  mechanism 
leaves its notable fingerprint on  
the flavor stability of a dense
neutrino gas   
by  coupling  flavor  conversion modes on  
different  scales 
which in turn, makes the flavor instabilities almost ubiquitous in the Fourier space.
This intriguing phenomenon arises from the fact that 
unlike the case of collective neutrino oscillations in a homogenous medium,
 the neutrino flavor conversion modes at different supernova zones depend linearly on
 each other   in a turbulent medium.
The most remarkable consequence of this effect is in that 
 it allows for  the presence of significant flavor conversions in the deepest supernova
 regions 
 even in the absence of the so-called fast modes. 
This is yet another crucial impact of turbulence on the physics of core-collapse
supernovae which can profoundly change our understanding 
 of neutrino flavor conversions in the supernova environment.

\end{abstract}

\maketitle

\section{Introduction}
Core-collapse supernova (CCSN) explosions are among the most energetic astrophysical phenomena
 in which neutrino emission is a major effect~\cite{Janka:2012wk, Burrows:2012ew}.
Neutrino flavor evolution  in CCSNe is a very
rich and nonlinear phenomenon in which neutrinos   
can experience  collective oscillations 
 due to the high density of the ambient neutrino 
 gas in the SN environment~\cite{Pastor:2002we,duan:2006an, duan:2006jv, duan:2010bg}. 
  In this paper, we study collective neutrino oscillations
  in the presence of SN turbulent matter density fluctuations which as discussed later herein,
  can significantly impact the physics of neutrino oscillations in CCSNe.

Collective neutrino oscillations could significantly impact the physics of CCSNe.
On the one hand, it could influence the SN dynamics and the nucleosynthesis 
of  heavy elements~\cite{Qian:1996xt} in the SN environment
 by modifying  the neutrino and antineutrino energy spectra and consequently, their interaction rates. 
On the other hand, understanding of collective neutrino oscillations is crucial
for future observations of galactic CCSNe neutrino signals~\cite{Scholberg:2012id, GalloRosso:2018ugl} 
and the upcoming measurements 
of diffuse supernova neutrino background~\cite{Beacom:2010kk}.

The first studies on collective neutrino oscillations in CCSNe were 
carried out in  maximally symmetric models, e.g.,
the stationary spherically symmetric \emph{neutrino bulb} 
model~\cite{duan:2006an, duan:2006jv, duan:2007sh, dasgupta:2009mg, duan:2010bg,
 Galais:2011gh, Duan:2007bt, Galais:2011gh, Duan:2015cqa}.
Within these simplistic models it was observed that
 the onset of collective neutrino oscillations can be at radii much smaller  than 
 that of the conversions induced by ordinary matter  via the 
 Mikheyev-Smirnov-Wolfenstein (MSW) mechanism (at least in  CCSNe with  iron cores).
 Despite this,
collective oscillations was still found to be suppressed in very  deep SN regions due to the presence of high
neutrino/matter densities~\cite{Duan:2010bf,Sarikas:2011am,Chakraborty:2011nf}. 
However, it was then realized  that in 
multidimensional (multi-D) time-dependent
 SN models, these suppressions can be dismissed thanks to 
 the breaking of  spatial/temporal symmetries~\cite{raffelt:2013rqa, duan:2013kba, duan:2014gfa,
 abbar:2015mca, Abbar:2015fwa, chakraborty:2015tfa, Chakraborty:2016yeg,
 Mirizzi:2015fva, Martin:2019kgi, Mangano:2014zda, Martin:2019dof}. 
 Yet,
in any realistic SN model, the physical conditions change so quickly that any unstable 
 mode  becomes stable before neutrinos can experience significant
flavor conversions~\cite{Chakraborty:2016yeg}. 
This means that in spite of the existence of flavor instabilities,
significant flavor conversions
 should be unlikely to occur in the 
deepest regions of the SN core.

Nevertheless, it was then perceived  that neutrinos can also experience the so-called \emph{fast}
flavor conversions on scales much shorter than those of traditional (slow) 
modes~\cite{Sawyer:2005jk, Sawyer:2015dsa,
 Chakraborty:2016lct, Izaguirre:2016gsx,  Wu:2017qpc,
  Capozzi:2017gqd, Richers:2019grc,  
   Abbar:2017pkh, Abbar:2018beu, Capozzi:2018clo,
 Martin:2019gxb, Capozzi:2019lso, Doring:2019axc, Johns:2019izj,
 Shalgar:2019qwg, Cherry:2019vkv, Chakraborty:2019wxe,  Abbar:2020fcl, Capozzi:2020kge, Xiong:2020ntn, Bhattacharyya:2020dhu,
 Shalgar:2020xns}. 
The fast scales
are determined by the neutrino number density, $n_\nu$, and can be as short as a few cm
in the deepest SN zones,
as opposed to the ones of slow modes which are determined by the vacuum 
frequency, $\omega = \Delta m^2/2E$, and
 occur on scales of $\sim$ a few km (for $\Delta m^2_{\rm{atm}}$ and  $E=10$ MeV neutrinos).
 Besides their phenomenological importance, perhaps the most remarkable physical consequence 
 of fast modes is in that they can lead to the occurrence of
 collective neutrino oscillations in the deepest regions of the SN core.
  This is because
 they occur on short enough scales in such a way that the unstable
 modes can experience significant flavor conversions before the physical conditions vary
 significantly. In spite of their importance, 
 fast modes do not seem to be
 a generic feature of CCSNe and even if they exist, they are thought to be 
 present only  in a finite region of the SN core~\cite{Abbar:2018shq, Abbar:2019zoq, 
 DelfanAzari:2019tez, Nagakura:2019sig, 
Morinaga:2019wsv,   Glas:2019ijo}. 
Additionally, fast modes may also be less likely to occur
in  non-exploding SN models~\cite{Abbar:2020qpi}.

 Turbulence plays a crucial  role in CCSNe~\cite{Abdikamalov:2014oba,
 Mabanta:2017kyb,Couch:2014kza,Radice:2017kmj,Meakin:2006uh}. 
 The impact of turbulent density fluctuations 
  on neutrino oscillations  has been extensively studied in 1D 
  models~\cite{Ma:2018key, Lund:2013uta, Reid:2011zz,Fogli:2006xy,
  Cherry:2011fm,Patton:2014lza,Yang:2015oya,Friedland:2006ta,
  Kneller:2007kg,Kneller:2010sc,Borriello:2013tha}, 
 where it can induce flavor conversions through parametric resonances. 
 Here, 
  we demonstrate that the presence of turbulence  
   in CCSNe can also induce \emph{collective} neutrino flavor conversion
  modes via an entirely different mechanism, i.e.,  the \emph{leakage} of flavor instabilities between different Fourier modes. 
  This novel effect
can significantly influence neutrino flavor evolution in  the SN environment and
in particular, it can lead to the presence of 
traditional (slow) collective neutrino oscillations
 in the deepest SN  regions 
 even in the absence of fast modes. 
What makes this novel effect
more promising is in that it  survives even for tiny  turbulence amplitudes.

\section{Linear Stability Analysis}
We 
start by deriving the equation of neutrino flavor evolution
in the linear regime, in the two-flavor scenario where
the flavor content of \textit{a} neutrino can be described as
\begin{align}
  \varrho = \frac{f_{\nu_e} + f_{\nu_x}}{2}
  + \frac{f_{\nu_e} - f_{\nu_x}}{2} \begin{bmatrix}
    s & S \\ S^* & -s \end{bmatrix},
\end{align}
where $f_\nu$'s are the neutrino initial occupation numbers 
 and, $S$
and $s$ carry information on neutrino  flavor coherence and conversion,
respectively. In the absence of collisions, the flavor evolution of the neutrino gas
 can be described by the 
Liouville-von Neumann 
equation ($c=\hbar=1$)~\cite{Sigl:1992fn,Strack:2005ux,Cardall:2007zw,Volpe:2013jgr, Vlasenko:2013fja}
\begin{equation}
i (\partial_t + \mathbf{v} \cdot \bm{\nabla})
\varrho_{E, \mathbf{v}} = \left[
 \frac{\mathsf{M}^2}{2E} + \frac{ \lambda}{2}\sigma_3 +
  \mathsf{H}_{\nu \nu, \mathbf{v}}, 
   \varrho_{E, \mathbf{v}}\right],
\label{eq:EOM1}
\end{equation}
 where  $\mathbf{v}$ is the neutrino velocity
   and $\lambda = \sqrt2 G_{\mathrm{F}} n_e$
  is the matter contribution to the neutrino Hamiltonian~\cite{Wolfenstein:1977ue,Mikheev:1986gs}. 
  Here, the energies and occupation numbers   
  are taken to be positive for neutrinos and negative for
antineutrinos,
   $\sigma_3$
  is the  third Pauli matrix and
  \begin{equation}
  \mathsf{H}_{\nu \nu, \mathbf{v}} = \sqrt2 G_{\mathrm{F}}
\int_{-\infty}^{\infty} \frac{E'^2\mathrm{d}E'}{(2\pi)^3} \int\!  \mathrm{d}\mathbf{v}'
  \varrho_{E',\mathbf{v}'}(1- \mathbf{v} \cdot \mathbf{v}')
\end{equation}
is the contribution from  the neutrino-neutrino forward scattering
\cite{Fuller:1987aa,Notzold:1988kx,Pantaleone:1992xh}.
  
We are here interested in the flavor stability analysis of neutrinos in the linear 
regime where the flavor conversion is still
insignificant and one has $s \simeq1$ and $|S| \ll 1$.
By only keeping terms of $\mathcal{O}(|S|)$ in Eq.~(\ref{eq:EOM1}),
one reaches~\cite{Banerjee:2011fj, Vaananen:2013qja}
\begin{equation}\label{eq:EOM}
i (\partial_t + \mathbf{v} \cdot \bm{\nabla}) S_{E,\mathbf{v}} = \big( \omega + \lambda + \Lambda_{\nu \nu, \mathbf{v}} \big) S_{E,\mathbf{v}}  - h_{\nu \nu, \mathbf{v}},
\end{equation}
where, with the definition $g_{E, \mathbf{v}} = 2\varrho_{E, \mathbf{v}}^{00}(t=0)$,
\begin{align}
 h_{\nu \nu, \mathbf{v}} &= \sqrt2 G_{\mathrm{F}}
  \int_{-\infty}^{\infty} \frac{E'^2\mathrm{d}E'}{(2\pi)^3}\! \int\!  \mathrm{d}\mathbf{v}'
  g_{E', \mathbf{v}'}  S_{E',\mathbf{v'}} ( 1- \mathbf{v} \cdot \mathbf{v}'), \nonumber\\
 \Lambda_{\nu \nu, \mathbf{v}} &= \sqrt2 G_{\mathrm{F}}
  \int_{-\infty}^{\infty} \frac{E'^2\mathrm{d}E'}{(2\pi)^3}\! \int\!  \mathrm{d}\mathbf{v}'
 g_{E', \mathbf{v}'}  ( 1- \mathbf{v} \cdot \mathbf{v}').
\end{align}

  Eq.~(\ref{eq:EOM}) provides a linear set of equations
   for which one can try  collective
  solutions of the form $S_{E, \mathbf{v}} = Q^{\Omega, \mathbf{k}}_{E, \mathbf{v}} 
  e^{-i\Omega t+i\mathbf{k\cdot x}}$ where $\Omega$ and $\mathbf{k}$
  satisfy the dispersion relation (DR) equation corresponding to Eq.~(\ref{eq:EOM}). 
  In a homogenous medium, this leads to 
  \begin{equation}\label{eq:lin}
 (-\Omega + \mathbf{v} \cdot \mathbf{k} + \omega + \lambda + \Lambda_{\nu \nu, \mathbf{v}})
  Q^{\Omega, \mathbf{k}}_{E, \mathbf{v}} =   h^{\Omega, \mathbf{k}}_{\nu \nu, \mathbf{v}}.
\end{equation}
Note that different Fourier modes are decoupled  
which means that $\mathbf{k}$ is just a \emph{parameter} here
and one only needs to find
 the functional form  of $Q^{\Omega, \mathbf{k}}_{E, \mathbf{v}}$ in the $E-\mathbf{v}$ space
 for a solution of DR equation.

\section{Turbulent Matter Fluctuations}
It 
simply follows from Eq.~(\ref{eq:lin}) that in a homogenous medium
where matter is constant, the matter potential $\lambda$ can be absorbed 
in the real part of $\Omega$ and therefore, does not  affect the stability 
condition of the dense neutrino gas.
However, if the matter is not  constant
and spatial density fluctuations are present, Eq.~(\ref{eq:lin}) changes to
\begin{equation}\label{eq:turb}
 (-\Omega + \mathbf{v} \cdot \mathbf{k} + \omega  + \Lambda_{\nu \nu, \mathbf{v}}) 
 Q_{E, \mathbf{v}}^{\Omega,\mathbf{k}} 
 +\int\! \frac{\mathrm{d}^3 \mathbf{k'}}{(2\pi)^3}\ \lambda_{\mathbf{k'}} Q_{E,\mathbf{v}}^{\Omega, \mathbf{k-k'}} =   
 h^{\Omega, \mathbf{k}}_{\nu \nu, \mathbf{v}},
\end{equation}
where $\lambda_{\mathbf{k}}$ is the Fourier component of the matter potential~\cite{Airen:2018nvp}.
Note that, most remarkably, different Fourier
modes are now  coupled
through the turbulence-induced convolution term and simple plane waves are not anymore
eigenvectors of Eq.~(\ref{eq:turb}). 
This  implies that  in order to solve Eq.~(\ref{eq:turb}),
one should  also consider the distribution of $Q_{E, \mathbf{v}}^{\Omega, \mathbf{k}}$ in the Fourier space
because $\mathbf{k}$ is not a parameter anymore
and  eigenvectors of  Eq.~(\ref{eq:turb}) can now have contributions from a range of  $\mathbf{k}$'s.

\begin{figure*} [t!]
 \centering
\begin{center}
\includegraphics*[width=.9\textwidth, trim=10 10 10 10, clip]{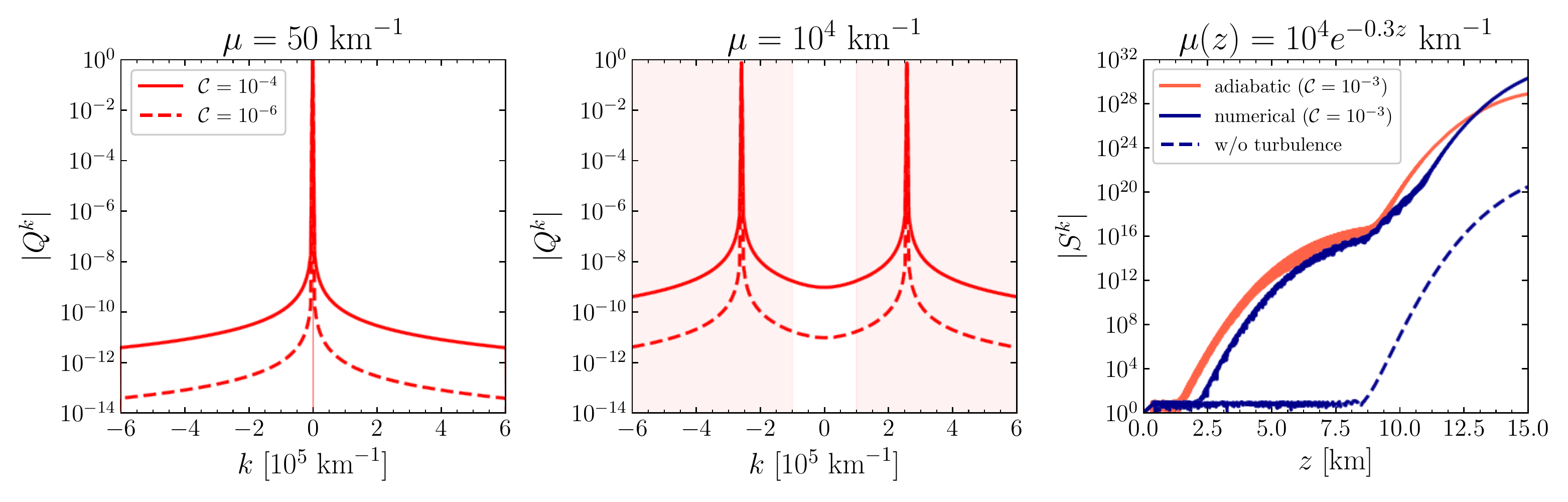}
\end{center}
\caption{
Left and middle panels: Overall shape of the eigenvectors of Eq.~(\ref{eq:turb}),
corresponding
to the  unstable mode  with the maximum growth 
rate for two representative values of $\mu$
and for
 $\mathcal{C} = 10^{-4}$  and $\mathcal{C} = 10^{-6}$.
 The shaded area indicates the unstable region in the homogenous 
 case (which is extremely narrow in the left panel). 
 Here, we have used Eq.~(\ref{eq:profile}) to relate matter to neutrino number density.
 Right panel: Evolution of the flavor coherence term in the linear 
 regime (Eq.~(\ref{eq:EOM})) for the representative $k = 10^3\ \mathrm{km}^{-1}$ Fourier mode
 in a declining $\mu$. Here,
 we  solve Eq.~(\ref{eq:EOM}) numerically for a discrete set of Fourier modes 
  with $\Delta k = 10^3$ km$^{-1}$ and assuming $\lambda_0=300\mu$.}
\label{fig:1}
\end{figure*}

In the following, we assume a Kolmogorov-like spectrum for  turbulence
where the matter density features power-law fluctuations~\cite{frisch1995turbulence}
on a range of scales between the dissipation scale, 
$\lambda_{\mathrm{diss}}$ ( here $\lambda_{\mathrm{diss}} \ll 10^{-10}$~km~\cite{Abdikamalov:2014oba}),
below which the turbulent energy gets efficiently dissipated by viscosity
and the cutoff scale,  $\lambda_{\mathrm{cut}}$, which is determined by the shock radius $R_\mathrm{s}$
so that $\lambda_{\mathrm{cut}} \sim 2 R_\mathrm{s}$.

 To be specific,
we take the turbulent matter fluctuations to have the form  
\begin{equation}\label{eq:Kolm}
\lambda(x) = \lambda_0\big( 1+ \mathcal{C} \sum_{k\neq0} \xi_k \cos(kx+\eta_k) \big),
\end{equation}
where  $\lambda_0$ is the average matter potential (the zeroth mode),
$\eta_k$ is a random phase and
$\mathcal{C} = \mathfrak{C}/ \mathfrak{C}_{\mathrm{N}}$
with $\mathfrak{C}$ and $\mathfrak{C}_{\mathrm{N}}$ being
 a constant coefficient and  a normalization
factor defined as $\mathfrak{C}_{\mathrm{N}} = (\sum_{k\neq0} \xi_k^2)^{1/2}$, respectively.
Here $\mathcal{C}$  is the most meaningful  parameter which specifies the
relative turbulence 
amplitude on scales $\sim \lambda_{\mathrm{cut}}$. In addition, $\xi_k$
is assumed to have a Kolmogorov distribution 
\begin{equation}
\xi_k = \bigg(\frac{k}{k_{\mathrm{cut}}}\bigg)^{-\alpha/2},
\end{equation}
with $k_{\mathrm{cut}} = 2\pi/\lambda_{\mathrm{cut}} $ which is fixed to
be $k_{\mathrm{cut}} = 0.01\ \mathrm{km}^{-1}$ in our calculations. 
We also set $\alpha = 5/3$ though our results 
do not depend qualitatively on the value of $\alpha$ for reasonable choices.
With these choices, one has $\lambda_k \sim \mathcal{C}\lambda_0 \big(k/k_{\mathrm{cut}}\big)^{-\alpha/2}$.

\section{Two-Beam Model}
We 
study neutrino flavor instabilities in a stationary 2D two-beam, monochromatic neutrino 
 gas studied first in Ref.~\cite{duan:2014gfa} (this stationary model is chosen for illustrative purposes, 
 otherwise see Appendix~\ref{temporal} for the  turbulence effect on temporal instabilities).
 Such a model can be used to describe the SN geometry  at some distance  from the 
 SN core~\cite{Chakraborty:2016yeg}
 where a periodic boundary condition is imposed in the transverse plane (along the $x$-axis in our model)
 and we study the evolution of the neutrino gas along the $z$-axis which can  also be
 interpreted as being the radial direction in spherical coordinate.
 The mono-energetic $\nu_e$ and $\bar{\nu}_e$ beams ($\omega = \pm1$)
 are assumed to be emitted with $\mathbf{v}_{\pm} = (\pm u,0,v_z)$
 where $u = \sqrt{1-v_z^2}$ with 
   $v_z=1/2$ (corresponding to
 an opening angle of $2\pi/3$ between the two beams)
 and the ratio between the number densities 
 is fixed to be $n_{\bar{\nu}_e}/n_{\nu_e} = 0.7$. 
 

We  solve Eq.~(\ref{eq:turb}) for our stationary model ($\Omega=0$) to find unstable modes
in the $z$-direction (where the imaginary part of $k_z$ is positive)
and the $E-\mathbf{v}$ distributions
of the corresponding eigenvectors 
as a function of the Fourier mode in the $x$-direction, $k$ (hereafter we drop the subscripts $x$
and superscripts $\Omega$).
In addition, for the small turbulence amplitudes  considered here
one can safely ignore the turbulence in the $z$ (longitudinal) direction  since
it is completely suppressed by the other terms in the equation of motion.
This implies that  Fourier modes are only coupled in the $x$ (transverse) direction.
Note that a similar suppression does not exist for the turbulence  in the $x$-direction
since there is no other term being able to compete  with 
the turbulence (coupling) effect.

To illustrate how  turbulent matter density fluctuations impact the stability of  
a dense neutrino medium,  
in Fig.~\ref{fig:1} we indicate the  overall shape of the eigenvectors of Eq.~(\ref{eq:turb}),
 defined as 
\begin{equation}
|Q^{k}| = \bigg(\sum_{E, \mathbf{v}} |Q_{E, \mathbf{v}}^{k}|^2 \bigg)^{1/2},
\end{equation}
corresponding
to the  unstable  mode  with the maximum growth rate,
where the eigenvectors are normalized to have unit length.
In the left panel, we first consider a neutrino gas with a relatively low neutrino
number density,  $\mu=\sqrt2G_{\mathrm{F}}n_{\nu_e}=50\ \mathrm{km}^{-1}$.
For such values of $\mu$, only very low Fourier modes are  unstable
in a homogenous neutrino gas. However, the instability structure 
changes dramatically in a turbulent medium.
As expected, there is a dominant peak with $|Q^k|\simeq 1$ at $k\simeq0$.
But due to the presence of  turbulent matter fluctuations, 
one can clearly see that the instability can now
\textit{leak} to  much higher Fourier modes (and make them unstable) which are otherwise completely  
stable in the homogenous case. It is of utmost importance to note that
in spite of its small amplitude (for the tiny turbulence amplitudes used here),
the leakage of instabilities can entirely 
change the stability condition of the neutrino gas.
This simply comes from the fact that
as far as the flavor stability  is concerned, the amplitude (of the unstable modes) is not relevant
since any unstable mode can growth exponentially with
a growth rate of $\sim 10$~km$^{-1}$ (for slow modes). 
Thus,  even unstable modes with amplitude  $|Q^k| \sim 10^{-9}$
can   get
activated within only $\sim 2$ km.
This implies
 that,  surprisingly, even a tiny  amount of turbulence in matter 
 would be enough to have a notable influence and make 
  the whole  range of  (relevant) Fourier modes
 unstable with reasonable initial amplitudes.
 This is very different from the turbulence-induced parametric resonances 
 where turbulent matter fluctuations can only generate a noticeable effect when
 the turbulence amplitude is  considerably large~\cite{Lund:2013uta}.   
 To the best of our  knowledge, the flavor instability leakage  is the only  
 physical effect in CCSNe which is sensitive to such tiny turbulence amplitudes.
 Indeed, the turbulence effect behaves here like a switch-on effect. 
Thus, one might be tempted to interpret the leakage phenomenon  
 as an example of the effect of the  \emph{background} symmetry breaking
in a dense neutrino gas.
Note  that in the absence of turbulence, $|Q^k|$  should be 
a $\delta$-function in the Fourier space.

The turbulence-induced leakage amplitude 
is  almost independent of $\mu$ and   depends  only on the  density
fluctuation amplitude (see Appendix~\ref{app1}) 
  \begin{equation}
\mathrm{leakage\ of}\quad k_0 \rightarrow k_0 \pm  k:\quad \frac{|Q^{k_0 \pm k}|}{|Q^{k_0}|} \sim \frac{\lambda_{k}}{k}.
\end{equation}
By using this expression, one can easily make an estimate of the leakage amplitude
for a given matter density and turbulence amplitude. 

In the middle panel of Fig.~\ref{fig:1}, an example of the instability leakage 
 for a high neutrino number density with $\mu=10^4$~km$^{-1}$ is presented.
For such a value of $\mu$ which is expected 
in the SN zones close to the surface of the PNS, only very large
$k$'s are unstable in the homogenous case. However, 
the instability  leaks to small $k$'s in the
presence of turbulence.
In particular, the leakage amplitude for a given turbulence amplitude
 is much larger in this case since the matter density is quite big in the 
 vicinity of the PNS.
 
 Although the form of the eigenvectors of Eq.~(\ref{eq:turb}) 
 changes significantly in a turbulent medium, its eigenvalues do not change
 noticeably, at least for such small turbulence amplitudes tried here.
 This implies that this novel effect observed for constant $\mu$'s
   might be still superficial unless it can also leave its influence on
    the solutions
   of Eq.~(\ref{eq:EOM}) for a realistic SN profile where $\mu$ is changing.
  But this is exactly where the power of the leakage mechanism is best manifested,
  as illustrated in the  right panel of Fig.~\ref{fig:1}. Here to provide a flavor of this effect, 
  we show
  the evolution of the $k=10^3$~km$^{-1}$ Fourier mode in a model in which
  the neutrino number density is  varying with
  $\mu(z) = 10^4 \mathrm{exp}(-0.3 z)$~km$^{-1}$ (note that $\mu$ 
  changes very rapidly and goes from $10^4$ to $10$~km$^{-1}$
  within only $\sim20$ km). 
  As can be clearly seen, the final amplitude of the 
  Fourier modes (at the point they become dominant) in the presence of turbulence
   can be larger than those of the homogenous gas by many orders of magnitude.
   This is due to the fact that all the relevant Fourier modes can always 
   grow exponentially in a turbulent medium in contrast to the homogenous gas in which each Fourier mode
   has a certain range of instability. This behavior is completely compatible with/predictable
    from what already observed in the left and middle panels of Fig.~\ref{fig:1}
    and shows that the assessment based on the shape of the eigenvectors of Eq.~(\ref{eq:turb})
     can be very useful in providing a sufficient insight on how Fourier modes grow.
     Note that the exact turbulence-induced enhancement in the amplitude
     of a  Fourier mode depends on the duration on which the turbulence 
     influences its evolution, which can be much longer for realistic SN profiles~\cite{unpublished2}. 
   
  The evolution of the neutrino gas here is adiabatic to a very good degree in the sense that
 the scales on which the eigenvectors of Eq.~(\ref{eq:turb}) grow (exponentially) are much shorter
  than those of variations in $\mu$ (or in the shape of the eigenvectors themselves), 
  i.e., $\kappa^{-1} \ll \mu/(\rm{d}\mu/\rm{d}r)$.
  One can then better understand the behavior observed in the right panel of Fig.~\ref{fig:1}
  in an analytical way, assuming a perfect adiabaticity.
  In the perfect adiabatic limit, the solution of Eq.~(\ref{eq:EOM}) at each  step
  $z=z_0+\Delta z$ can be obtained analytically from the one at $z=z_0$ by 
  $S(z_0+\Delta z) = \sum_i c_i \Psi_i e^{i (k_z)_i \Delta z} $
  where $\Psi_i$ and $(k_z)_i$ are the eigenvectors 
  (which form a complete basis) and eigenvalues 
  of the Hamiltonian of Eq.~(\ref{eq:turb})
  at  $z=z_0$ and $c_i$'s are the expansion coefficients of $S(z_0) = \sum_i c_i \Psi_i$.
  Such an analytical  adiabatic solution (red curve)  shows a very good agreement with the
  numerical solution of Eq.~(\ref{eq:EOM}).  
  One should note that 
the key point here is  that the eigenvectors of Eq.~(\ref{eq:turb})
  at two different  steps \emph{are not exactly linearly independent} of each other.
  In other words, each $\Psi_i^{\rm{new}}$ has contributions from all $\Psi_j^{\rm{old}}$'s, i.e.,
  $\Psi_i^{\rm{new}} = \sum_j c_{ij} \Psi_j^{\rm{old}}$ with 
   $c_{ij}$ being  rouphly  the turbulence amplitude.
   This means that any unstable mode grows from an enhanced 
    initial value  
    (occurred during the growth of the modes which were  previously unstable)
    which in turn ensures that all the Fourier modes always grow exponentially 
   during the propagation of neutrinos.
  This is entirely in contrast to the homogenous case where the new unstable 
  modes at each point are totally linearly independent of the old ones
  at a previous point and therefore, any exponential growth is present only within a certain period.


\section{Discussion and Conclusions}
The 
turbulence-induced leakage of flavor instabilities 
implies that  
the notion of  $\mu-k$ instability 
band (see, e.g., Fig.~3 in Ref.~\cite{abbar:2015mca})
 developed in a homogenous neutrino gas is not very useful
in a turbulent medium where what distinguishes different Fourier modes
is actually only their initial amplitudes, $|Q^k|$.

\begin{figure} [tb!]
 \centering
\begin{center}
\includegraphics*[width=.45\textwidth, trim= 10 10 10 10, clip]{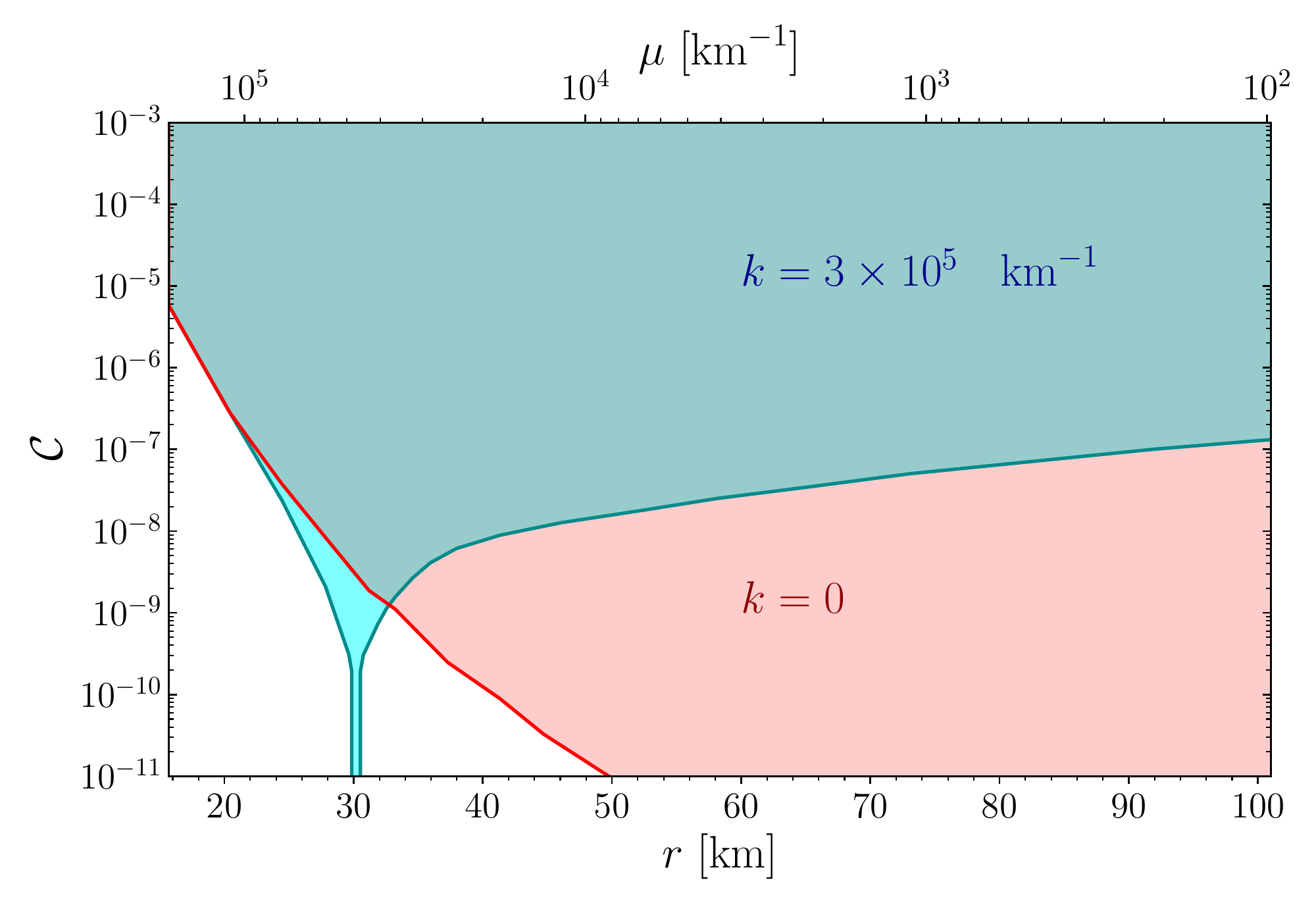}
\end{center}
\caption{Instability regions (shaded areas) of two representative Fourier modes, $k=0$
and $3\times10^5\ \mathrm{km}^{-1}$, 
as a function of  the distance from the SN core   and the turbulence amplitude.
Here we have employed the
two-beam model described in the text.
}
\label{fig:2}
\end{figure}

One can immediately observe that  the leakage mechanism can well address  
one of the biggest issues with slow instabilities in the deepest 
SN zones.
In particular, it
 dismisses the necessity of the occurrence of fast modes 
in order to observe significant flavor conversions near the PNS.
To demonstrate this idea, in Fig.~\ref{fig:2} we show  the instability footprints
 of two representative Fourier modes
   as a function of the distance
 from the SN core and the turbulence amplitude, $\mathcal{C}$,
 during the  accretion phase of a CCSN\footnote{ Here we  \emph{define} the unstable region
for a given Fourier mode  by requiring $|Q^k|>10^{-13}$ for  
the eigenvector of the mode with maximum growth rate. This is to ensure that 
the activation scale of a given mode is shorter than the variation scales of $\mu$. 
Otherwise,  such boundaries for the unstable regions are absolutely artificial.}.
 Here we take a matter/neutrino
 density profile approximately similar to that of Ref.~\cite{Chakraborty:2016yeg}  in which 
 \begin{equation}\label{eq:profile}
 \mu(r) = \mu_R (R/r)^4 \quad   \mathrm{and}\quad \lambda(r) = \lambda_R (R/r)^3,
 \end{equation}
where 
$\mu_R$ and $\lambda_R$ are the neutrino and matter densities
 on the surface of the neutrinosphere, $R$, respectively,
for which we have used $R=15$~km, $\mu_R = 2\times10^5\ \mathrm{km}^{-1}$
and $\lambda_R = 6\times10^7\ \mathrm{km}^{-1}$ (corresponding
to a matter density of $\rho\simeq 3\times 10^{11}\ \mathrm{g\ cm}^{-3}$).
For very small turbulence amplitudes, the instability zones 
can be  extremely narrow specially at small radii
which prevents any significant flavor conversions therein.
However,
as $\mathcal{C}$ increases, the Fourier modes  become unstable
at all radii which means that they can grow many orders of magnitude (as in the right panel of Fig.~\ref{fig:1})
and easily enter the nonlinear regime.
Hence, \emph{no matter} whether fast modes exist or not, 
collective neutrino oscillations can occur within just a few km above the PNS.
In addition, unlike  fast modes which can only exist in small SN
regions and are less likely to occur in non-exploding models, 
turbulence-induced flavor conversion modes are ubiquitous and generic.
This could  have an important impact
 on the SN dynamics
and the nucleosynthesis of heavy elements in CCSNe.


Apart from the crucial impact of  turbulence on the flavor stability of a dense neutrino gas,
its presence is also  important in providing 
 initial seeds for the unstable modes. Specifically, the turbulence term
  in Eq.~{\ref{eq:EOM}} transfers the initial seed from  $k_0$ to
  $k_0 \pm k$ on scales $\sim \lambda_{ k}^{-1}$, or more accurately, 
  $\sim \mathrm{max}\{\lambda_{k'}^{-1} k/ k'\}$ where the maximum is taken over 
  all turbulence modes. 
 
Apart from its impact on slow modes which was discussed here,  
turbulence can also affect fast neutrino flavor conversion modes (see Appendix~\ref{fast}).
However, although a similar leakage  can occur therein, the leakage mechanism
does not  seem to change   DR of fast modes.

In the above discussions, we have only considered the effects of  spatial 
density fluctuations  on the spatial instabilities. However, 
 a similar
effect should  also be  expected for temporal instabilities as shown in Appendix~\ref{temporal}. 
Indeed, the leakage effect has nothing to do with the eigenvalues of DR equation and  the nature of instabilities
 and, it only arises due to the presence
of coupling among different  eigenvectors. 
Additionally as discussed in Appendix~\ref{fast},
such an effect even exists for stable solutions (real eigenvalues of DR equation).
Similarly,  temporal fluctuations of the matter density
can also couple different temporal frequencies. Although extremely
 rapid temporal density variations are necessary to observe any noticeable effect,
 it could be still plausible considering the small required 
 turbulence amplitudes.
Moreover, while we have only considered the effects of the turbulence on flavor instability
in  CCSNe, similar effect can be expected in any dense neutrino
environment where matter density fluctuations are present, such as
 neutron star merger remnant accretion disks.

Our study is meant only to serve as an introduction to this novel issue 
and further research is necessary to provide a better understanding
of its physical implications. This is yet another time that the rich physics of neutrino flavor
evolution in dense neutrino media surprises us.



\section*{Acknowledgments}
I am  deeply indebted to Georg Raffelt and Huaiyu Duan for their comments on the
manuscript, their kind encouragements and many insightful discussions during the 
development of this work. 
I am also  grateful to
  Cristina Volpe, Ernazar Abdikamalov and Shashank Shalgar for useful conversations/communications
 and to Hans-Thomas Janka and Francesco Capozzi for reading the manuscript and their comments.
I  acknowledge partial support by the Deutsche 
Forschungsgemeinschaft (DFG) through Grant No. SFB 1258 (Collaborative Research Center Neutrinos, Dark Matter, Messengers).





\appendix
\section{The Leakage Amplitude}\label{app1}
We here  develop a hand-waving understanding  of the turbulence-induced leakage
amplitude. 
We  
take the  structure of the equation of  neutrino flavor evolution
in the linear regime and attempt
to understand how the presence of a coupling between different Fourier modes
changes the structure of the eigenvectors.
For this purpose,  we consider the following  set of equations which resembles the evolution
of the flavor coherence   terms of two  Fourier modes which are coupled 
\begin{align}
i\partial_t S^{k_1} &= (\omega+\mu+k_1)S^{k_1} + \lambda_k S^{k_2}\nonumber\\
i\partial_t S^{k_2} &= (\omega+\mu+k_2)S^{k_2} + \lambda_k S^{k_1}
\end{align}
where $k=k_2-k_1$. It can be easily shown that the eigenvectors of this
 set of linear equations have the form $|Q| \propto (1,\lambda_k/k)$ and
$(\lambda_k/k, 1)$ for $\lambda_k/k\ll1$ (which is always the case for the
Fourier modes of interest). This confirms that in the presence of a coupling
term, any eigenvector will have a dominant component at a given Fourier mode
and  subdominant contributions from  other modes with  amplitude
$\sim \lambda_k/k$.

\section{temporal instabilities}\label{temporal}
We have studied the leakage mechanism
in a stationary dense neutrino gas. Here, we demonstrate that 
a very similar effect arises in a time-dependent model where
 turbulence can impact the temporal instabilities. This  should be  of course
expected since the leakage effect has nothing to do with the eigenvalues of DR equation and  
the nature of instabilities
 and, it only arises due to the presence
of coupling among different  eigenvectors. 
Hence, as will also be discussed in our upcoming work~\cite{unpublished2},
the leakage effect  exists also for multi-angle configurations.

We here consider a time-dependent  two-beam, monochromatic neutrino gas
with one spatial dimension. Our model is the same as the one proposed in Ref.~\cite{Martin:2019gxb}  
but we only consider two (zenith) angle beams  with emission angles 
$\vartheta_1=\pi/6$ and $\vartheta_2=\pi/3$ with respect to the $z$-axis.
We here  assume that
the neutrino gas possesses a perfect axial symmetry about
the $z$-axis.

\begin{figure} [thb!]
 \centering
\begin{center}
\includegraphics*[width=.37\textwidth, trim= 0 10 0 10, clip]{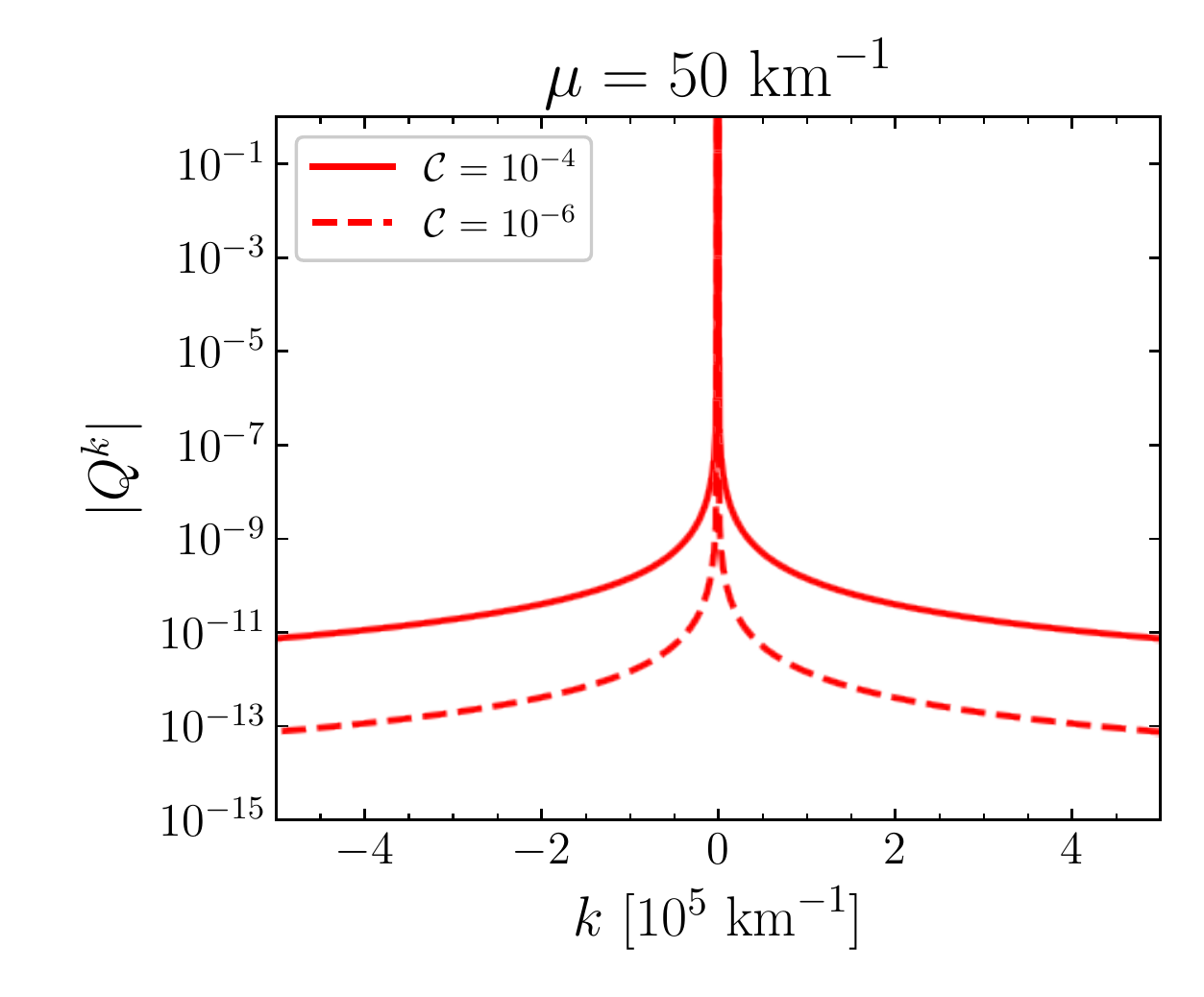}
\end{center}
\caption{
Overall shape of the eigenvectors (of Eq.~(7) in the main text) for the time-dependent neutrino
gas described here,
corresponding
to the  unstable \emph{temporal} mode  with the maximum growth 
rate for  $\mu=50$~km$^{-1}$.
 Here to relate matter to neutrino number density we have used Eq.~(12) in the main text.
}
\label{fig:temp}
\end{figure}

The results obtained in this model are presented in Fig.~\ref{fig:temp}
where we have as an example shown the  overall shape of $|Q^k|$ (here k is the Fourier mode in $z$-direction) 
for $\mu=50$,
corresponding to the unstable \emph{temporal} mode with the maximum growth rate.
As can be obviously seen, the temporal instabilities are  similarly
affected by the leakage mechanism.

\section{Fast modes}\label{fast}

\begin{figure} [thb!]
 \centering
\begin{center}
\includegraphics*[width=.37\textwidth, trim= 0 0 15 35, clip]{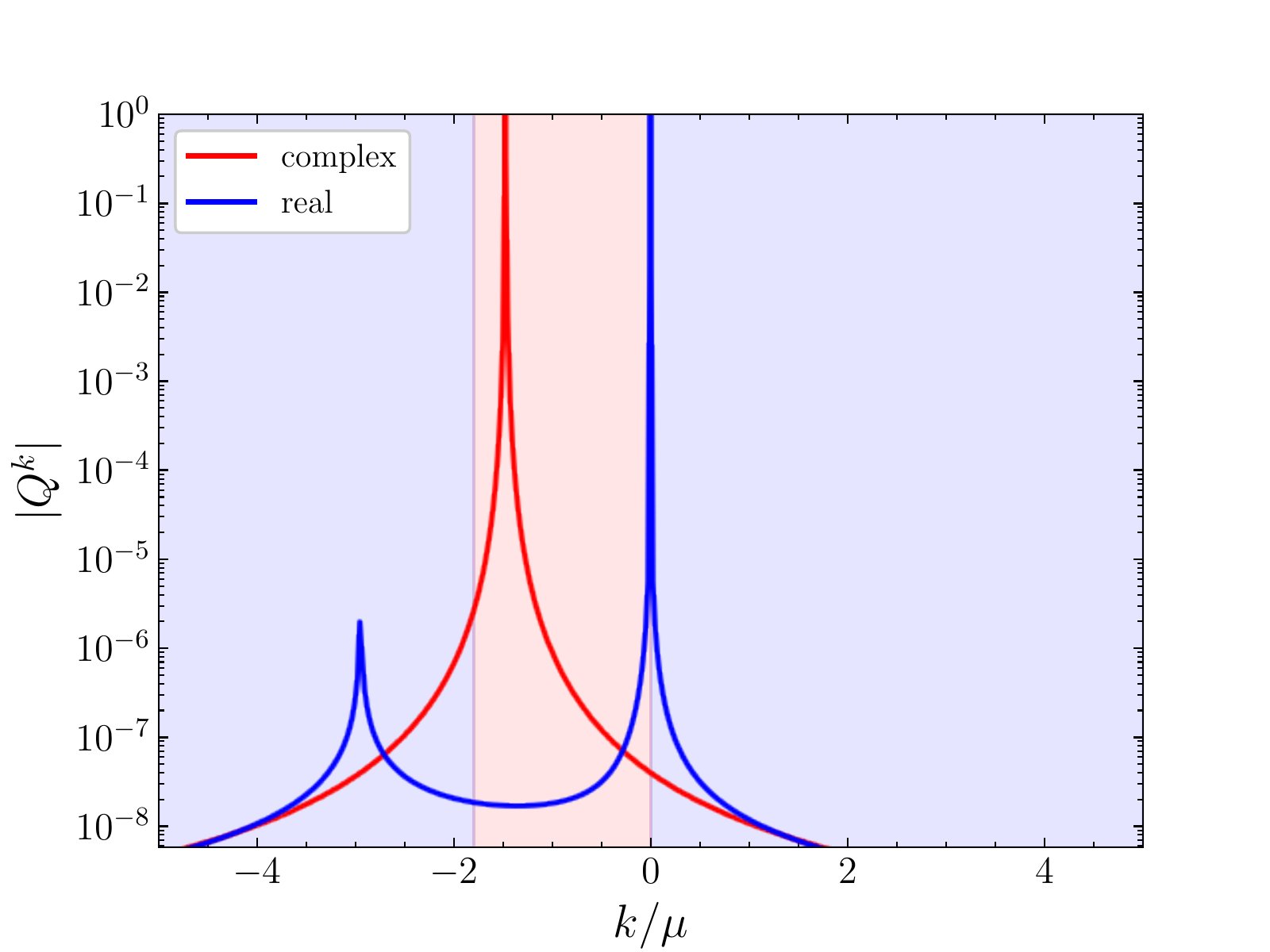}
\end{center}
\caption{
Overall shape of two arbitrary real and complex solutions (of
  Eq.~(7) in the main text), in the presence of fast modes
  in a turbulent medium, with $\mathcal{C}=10^{-6}$ and assuming $\lambda=300\mu$. The blue and red
areas show the regions where the real branches and the gap (the complex branches)
 exist in a homogenous neutrino gas, respectively. }
\label{fig:3}
\end{figure}

Turbulent matter fluctuations can also  influence
 fast modes in a similar way to slow modes.
This  is clearly illustrated in Fig.~\ref{fig:3} where
two arbitrary stable and unstable modes are shown
for a neutrino gas in the presence of fast modes.
Here we have  
considered a 2D stationary  neutrino gas  with only one neutrino and
one antineutrino beams with 
$\mathbf{v}_{\bar\nu_e} = ( +u,0,v_z)$ and
$\mathbf{v}_{\nu_e} = ( -u,0,v_z)$ 
 ($v_z=1/2$)
 in such a way that fast modes can exist (see
 the model studied in Ref.~\cite{Izaguirre:2016gsx}).
The blue and red
bands indicate the regions where the real branches and the gap (where
 complex branches exist) are located, respectively, in a homogenous neutrino gas. 
 In the presence of turbulence, 
both the real and complex solutions
can leak to the other zone.  Note  that  the  leakage phenomenon is not unique to the unstable
modes (complex branches) and stable modes also leak to the unstable region
(with $|Q^k|\sim \lambda_k/k$). 
Note, however, that considering the locality of fast modes, as long as the eigenvalues dispersion relation (DR) equation
  are not modified by the turbulence one can always define a new set of basis
  so that the DR remains unchanged. 


\bibliographystyle{elsarticle-num}
\bibliography{bib}


\end{document}